\documentstyle[graphics,twocolumn]{mn2e}

\newcommand{\mincir}{\raise
-2.truept\hbox{\rlap{\hbox{$\sim$}}\raise5.truept 
\hbox{$<$}\ }}
\newcommand{\magcir}{\raise
-2.truept\hbox{\rlap{\hbox{$\sim$}}\raise5.truept
\hbox{$>$}\ }}
\newcommand{\minmag}{\raise-2.truept\hbox{\rlap{\hbox{$<$}}\raise
6.truept\hbox
{$>$}\ }}

\newcommand{\ns}{{\mbox{n$_{\rm s}$}}}
\newcommand{\as}{{\mbox{A$_{\rm s}$}}}
\newcommand{\nt}{{\mbox{n$_{\rm t}$}}}
\newcommand{\nrun}{{\mbox{n$_{\rm run}$}}}
\newcommand{\omb}{{\mbox{$\Omega_{\rm b}$}}}

\newcommand{\beq}{\begin{equation}}
\newcommand{\eeq}{\end{equation}}
\newcommand{\teff}{\tau_{\rm eff}}

\newcommand{\lya}{Lyman-$\alpha$~}

\newcommand{\Mp}{M_{\rm Pl}}

\newcommand{\be}{\begin{equation}}
\newcommand{\ee}{\end{equation}}
\newcommand{\ba}{\begin{eqnarray}}
\newcommand{\ea}{\end{eqnarray}}
\newcommand{\brr}{\begin{array}}
 
\newcommand{\err}{\end{array}}
\newcommand{\bc}{\begin{center}}
\newcommand{\ec}{\end{center}}

\newcommand{\apj}{\mbox ApJ}

\newif\ifAMStwofonts
\AMStwofontstrue

\DeclareMathAlphabet{\mathsc}{OT1}{cmr}{m}{sc}
\def\testbx{bx}%
\DeclareRobustCommand{\ion}[2]{%
\relax\ifmmode
\ifx\testbx\f@series
{\mathbf{#1\,\mathsc{#2}}}\else
{\mathrm{#1\,\mathsc{#2}}}\fi
\else\textup{#1\,{\mdseries\textsc{#2}}}%
\fi}

\title[Constraints on the Primordial Power Spectrum from the
Ly$-\alpha$ Forest and WMAP]
{Constraints on the Primordial Power Spectrum from High Resolution
  Lyman$-\alpha$ Forest Spectra and WMAP} 
\author[M. Viel, J. Weller \& M.G. Haehnelt] {Matteo Viel,
Jochen Weller \& Martin G. Haehnelt$^{1}$ \\ $^1$ Institute of
Astronomy, Madingley Road, Cambridge CB3 0HA}

\begin{document}

\maketitle
\begin{abstract}
The combined analysis of the cosmic microwave background on large
scales and \lya forest on small scales provides a sufficiently long
lever arm to obtain strong constraints on the slope and curvature of
the power spectrum of primordial density fluctuations.  We present
results from the combination of the first year WMAP data and the dark
matter power spectrum inferred by Viel et al. (2004) for two different
sets of high resolution and high signal-to-noise quasar absorption
spectra: the Croft et al. (2002) sample with a median redshift $z=2.72$
and the LUQAS sample (Kim et al. 2004) with a median redshift
$z=2.125$. The best fit value for the {\it rms} fluctuation amplitude
of matter fluctuations is $\sigma_{8} =0.94 \pm 0.08$ and $\ns=0.99 \pm
0.03$, if we do not include running of the spectral index. The best fit
model with a running spectral index has parameters $\ns=0.959 \pm
0.036$ and $\nrun=-0.033\pm 0.025$.  The data is thus consistent with a
scale-free primordial power spectrum with no running of the spectral
index. We further include tensor modes and constrain the slow-roll
parameters of inflation.
\end{abstract}


\section{Introduction}

The Wilkinson Microwave Anisotropy Probe team (WMAP, Bennet et al. 2003,
Spergel et al. 2003) has presented impressive confirmation for what is
now considered the standard cosmological model: a flat universe
composed of matter, baryons and vacuum energy with a nearly
scale-invariant spectrum of primordial fluctuations.  One of the
surprising results of the WMAP team's combined analysis of cosmic
microwave background (CMB), galaxy redshift survey and \lya forest
data was the evidence for a curvature of the 
power spectrum of primordial density fluctuations.  This was expressed
by the WMAP team in terms of a non-vanishing derivative $\nrun= d \ns/
d\ln k$ of the spectral index $\ns$.  In inflationary models the shape
of the primordial power spectrum is directly related to the potential
of the inflaton field. For slow-roll inflation simple relations exist
between the slow-roll parameters and the ratio of tensor to scalar
fluctuation, the power-law index and the derivative of the power-law
index of primordial fluctuations (see Liddle and Lyth 2000 for a
review).  Constraints on the power spectrum of primordial density
fluctuations can thus at least in principle constrain inflationary
models (Hannestad et al. 2002, Peiris et al. 2003, Leach \& Liddle 2003, Kinney et al. 2004, Tegmark et al. 2004).

The \lya forest data constrains the dark matter (DM) power spectrum on
scales of $\sim 4-40 $ Mpc$/h$ in the redshift range $2<z<4$ (Croft et
al. 1998, Hui 1999, Croft et al. 1999, McDonald et al. 2000, Hui et
al. 2001, Croft et al. 2002 [C02], McDonald 2003, Viel et al. 2003,
Viel, Haehnelt \& Springel 2004 [VHS]).  Due to the long lever arm
between \lya forest on small scales and the CMB on large scales the
combined analysis of these data sets alone can provide strong
constraints on the slope and curvature of the primordial power spectrum in
the context of CDM-like models.  Croft et al. (1999) inferred an
amplitude and slope which was consistent with a COBE normalised
$\Lambda$CDM model with a primordial scale invariant fluctuation
spectrum (Phillips et al. 2001).  McDonald et al. (2000) and C02, using
a larger sample of better quality data, found a somewhat shallower
slope and smaller fluctuation amplitude.  This latter data was part of
the WMAP team's analysis mentioned above which gave evidence for a
tilted primordial CMB-normalised fluctuation spectrum ($\ns<1$) and/or
a running spectral index (Bennet et al. 2003; Spergel et al. 2003;
Verde et al. 2003).  There are, however, a number of systematic
uncertainties in the amplitude of the DM power spectrum inferred form
the \lya forest data (C02, Zaldarriaga, Scoccimarro \& Hui, 2003;
Zaldarriaga, Hui \& Tegmark, 2001; Gnedin \& Hamilton 2002; Seljak,
McDonald \& Makarov 2003, VHS).

VHS have recently used the method developed by C02 together with a
large suite of new high-resolution numerical hydro-simulations to
obtain an improved estimate of the DM power spectrum from the \lya
forest.  Two large samples of high-resolution spectra were used for
this analysis, the sample of HIRES/LRIS spectra in the original
analysis of C02 and the Large Sample of UVES QSO Absorption Spectra
(LUQAS) of Kim et al. (2004).  VHS demonstrated that the inferred {\it
rms} fluctuation amplitude of the matter density is about 20\% higher
than that inferred by C02 if the best possible estimate for the
effective optical depth, as obtained from high-resolution absorption
spectra, is used.  VHS further found  that  with this effective optical
depth the inferred DM spectrum is
consistent with a scale invariant COBE normalized DM power spectrum
thus confirming the findings of Seljak et al. (2003).
The running spectral index model advocated by the WMAP team falls
significantly below the DM spectrum inferred from the \lya forest data
(VHS).

In this paper, we will perform a combined analysis of the results of
VHS based on the two independent data sets of C02 and K04 together
with the WMAP data. The plan of the paper is as follows. In Section
\ref{data} we present the two data sets and briefly outline the method
used to infer the linear dark matter power spectrum. Section
\ref{analysis} contains a description of the method used to estimate
{\it rms} fluctuation amplitude $\sigma_8$, the spectral index $n_s$,   
a possible running of the spectral index and the parameters of  
slow-roll inflationary models.  In Section \ref{conclu} we summarise 
and discuss our results.

\section{The WMAP and \lya forest data}
\label{data}
\subsection{The WMAP data}
The CMB power spectrum has been measured by the WMAP team over a large
range of multipoles ($l<800$) to unprecedented precision on the full sky
(Hinshaw et al. 2003, Kogut et al. 2003). For the analysis we use the
one year release of 
the WMAP temperature and temperature-polarization cross-correlation
power spectrum\footnote{http://lambda.gfsc.nasa.gov}. The CMB power
spectrum depends on the underlying primordial fluctuation via the
photon transfer functions. The angular power spectrum in the
temperature exhibits acoustic peaks at $l=220$ with an amplitude of
$\Delta T \approx 75\, \mu {\rm K}$ and $l=540$ with $\Delta T \approx
49 \, \mu{\rm K}$ (Page et al. 2003).  The WMAP team (Spergel et
al. 2003) showed 
that the measured CMB power spectrum is in excellent agreement with a
flat universe and throughout this paper we will adopt this as a
prior. In order to estimate parameters from the CMB we exploit for
COSMOMC\footnote{http://www.cosmologist.info} an adapted version of the
CMB likelihood code provided by the WMAP team (Verde et al. 2003,
Hinshaw et al. 2003, Kogut et al. 2003).

\subsection{The \lya forest data}
The power spectrum of the observed flux in high-resolution \lya forest data 
constrains the DM power spectrum on scales of $0.003\,{\rm s/km} 
< k < 0.03\,{\rm s/km}$.  At larger scales the errors due to
uncertainties  in fitting a  continuum to the absorption spectra
become large while at smaller scales the contribution of metal
absorption systems becomes dominant (see e.g. Kim et al. (2004) 
and McDonald et al. (2004) for discussions). We will use here the DM 
power spectrum which VHS inferred for the 
C02 sample (Croft et al. 2002) and the LUQAS sample (Kim et al. 2004).  
The C02 sample consist of 30 Keck HIRES spectra and 23 Keck LRIS
spectra and has a median redshift of $z=2.72$.  The LUQAS sample 
contains 27 spectra taken with the
UVES spectrograph and has a median redshift of $z=2.125$. 
The resolution of the spectra is 6 km/s, 8 km/s 
and 130 km/s  for the UVES, HIRES and LRIS spectra, respectively. 
The S/N per resolution element is typically 30-50. 

The method used to infer the dark matter power spectrum from the observed flux has been
proposed by Croft et al. (1998) and has been improved by C02, Gnedin \&
Hamilton (2002) and VHS.  The method relies on hydro-dynamical
simulations to calibrate a bias function between flux and
matter power spectrum: $ P_F(k) = b^2(k)\;P(k)$, 
on the range of scales of interest, and has been found to be robust
(Gnedin \& Hamilton 2002).

There is a number of systematic uncertainties and statistical errors
which affect the inferred power spectrum and an extensive discussion
can be found in C02, Gnedin \& Hamilton (2002); Zaldarriaga, Scoccimarro
\& Hui (2003) and VHS. VHS estimated the uncertainty of the {\it
rms} fluctuation amplitude of matter fluctuation to be 14.5 \%. There
is a wide range of factors contributing to this uncertainty.  
In the following we present a brief summary. The effective optical
depth which is used to calibrate numerical simulation and has to be
determined separately from the absorption spectra is a major
systematic uncertainty. Unfortunately, as discussed in VHS there is a
considerable spread in the measurement of the effective optical depth
in the literature.  Determinations from low-resolution low S/N spectra
give systematically higher values than high-resolution high S/N
spectra. There is little doubt that the lower values from
high-resolution high S/N spectra are appropriate and the range
suggested in VHS leads to a 8\% uncertainty in the {\it rms}
fluctuation amplitude of the matter density field (Table 5 in VHS).
Other uncertainties are the slope and normalization of the
temperature-density relation of the absorbing gas which is usually
parametrised as $T=T_0\,(1+\delta)^{\gamma-1}$.  $T_0$ and $\gamma$
together contribute up to 5\% to the error of the inferred fluctuation
amplitude.  Our estimate of the uncertainties due to the method was
also 5\%.  Somewhat surprisingly VHS found substantial differences
between numerical simulations of different authors and assigned a
corresponding uncertainty of 8\%.   Other possible systematic errors 
include fluctuations of the UV background, temperature fluctuation in
the IGM due to late helium reionization and the effect of galactic
on the surrounding IGM. In VHS we had assigned 
an uncertainty of  5\% for these. Summed in quadrature this led to
the estimate of the overall uncertainty of 14.5\% in the {\it rms}
fluctuation amplitude of the matter density field.

For our analysis we take the inferred DM 
power spectrum in the range 
$0.003\,{\rm s/km} < k < 0.03\,{\rm s/km}$ 
as given in  Table 4 of  VHS. 
Note that we have reduced the values  by 7\%  to mimick 
a model with $\gamma=1.3$, 
the middle of the plausible range for $\gamma$ 
(Ricotti et al. 2000, Schaye et al. 2000).
In the next section we will also show results for the 
DM power spectrum inferred by C02 in their original analysis 
(their Table 3).

\section{Constraining the Primordial Power Spectrum and Inflation}\label{analysis}
\subsection{The Power Spectrum}
\label{powerspectrum}
We will first attempt to constrain the primordial power spectrum of
scalar fluctuations. We parametrise the power spectrum by
\beq
P(k) = P(k_0)\left(\frac{k}{k_0}\right)^{n_s(k)}\, ,
\eeq
where we choose the pivot scale to be $k_0 = 0.05\,{\rm Mpc}^{-1}$ for
the analysis in this subsection. The simplest case is a scale
invariant power spectrum with $n_s =1 $ constant.
In a more general context, and assuming $d^2\ns/d\ln k^2 =0$, the spectral index is defined by:
\beq
n_s(k) = n_s(k_0) + \frac{d\,n_s}{d\,\ln k}\ln\left(\frac{k}{k_0}\right)\;,
\eeq
with $\nrun \equiv d\,n_s / d\,\ln k$ parametrising the evolution of
the spectral index to first order.  
\begin{figure*}
\center\resizebox{1.0\textwidth}{!}{\includegraphics{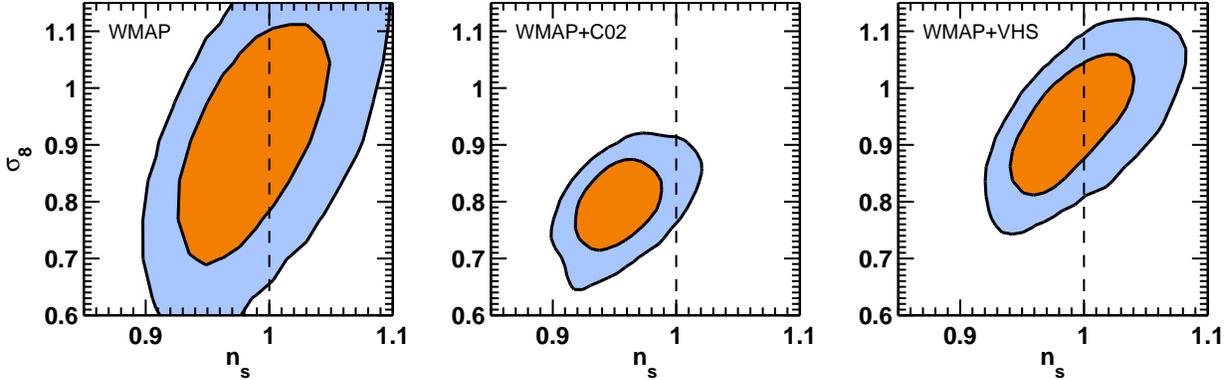}}
\caption{Joint $1$- and $2$-$\sigma$ likelihoods on $\sigma_8$ and
$\ns$. In the left panel the constraints from WMAP only, the middle
panel from WMAP combined with the \lya~power spectrum of C02, for a $\teff=0.349$.
On the right the constraints from WMAP combined
with VHS, for a $\teff (z=2.125) = 0.17$
and $\teff (z=2.72) = 0.305$. The dashed lines indicate the spectral index
for a scale invariant $\ns=1$ primordial spectrum.}
\label{fig:norun}
\end{figure*}
We extended the COSMOMC package (Lewis \& Bridle 2002) to
include the \lya data as discussed in Section \ref{data}. As mentioned
before the main uncertainty of the
\lya data is its relative calibration with respect to the underlying
dark matter power spectrum. This is mainly influenced by the effective
optical depth $\tau_{\rm eff}$,  the normalization and slope of the
temperature-density relation, $T_0$ and $\gamma$ and the other
uncertainties described in section 2.2. We describe this
uncertainty with an effective calibration amplitude $A$ and obtain for the
log-likelihood
\beq
-\ln {\cal L}_{Ly\alpha} \propto \sum_k\frac{\left[A\,P_{\rm dat}(k)-P_{\rm
th}(k)\right]^2}{\sigma_P^2}\, .
\eeq
We introduce the calibration $A$ as an extra parameter into the Markov
chain and multiply the likelihood with the prior
\beq
P_{\rm prior} \propto
\exp\left[-\frac{1}{2}\frac{\left(A-{\bar A} \right)^2}{\sigma_A^2}\right]\, .
\eeq
In order to obtain constraints on the cosmological parameters we will
marginalize over the calibration parameter $A$. Note that this differs
from the method described in Verde et al. (2003).
As a standard prior on the mean we choose ${\bar A}=1$.

The cosmological parameters which are varied for the combined analysis
are the physical baryon density $\omb\, h^2$, the physical dark
matter density $\Omega_{\rm c}\, h^2$, the angular diameter distance to last
scattering $\theta$, the optical depth $\tau$ due to instantaneous complete
reionization, the amplitude $\as$ of primordial perturbations and as
mentioned above the spectral index of scalar perturbations, either
running or constant. Note that we restrict the analysis to a flat
universe with a cosmological constant, where we neglect the
contribution from neutrinos and tensor modes. We further restrict the
analysis to optical depths $\tau < 0.3$ as suggested by the WMAP team
(Spergel et al. 2003, Verde et al. 2003). The amplitude of the matter
fluctuations in a 8 Mpc$/h$ sphere, $\sigma_8$, is a derived quantity
for this analysis.

We first discuss the results when we assume a constant spectral index
$\ns$ ($\nrun =0$).

In Figure~\ref{fig:norun} we show the joint
likelihood for the scalar spectral index $n_s$ and $\sigma_8$. On the
left is the result for WMAP only. Note that we {\em only} include the
WMAP data set and {\em not} the extended CMB data set in order to
concentrate on the constraints which are obtained by adding the
\lya~forest.  The  WMAP data alone gives only very weak limits 
on  the amplitude of the matter fluctuations and the spectral index. 
For a marginalized likelihood the spectral index is $n_s = 0.990 \pm
0.039$, consistent with a scale
invariant power spectrum as found by the WMAP team and other authors.
In the middle panel we show the result if we combine the WMAP data
with \lya data from C02 as compiled by
Gnedin \& Hamilton (2002). Note that we impose an uncertainty on the
calibration of $\sigma_A = 0.25$ if $A<1$ and $\sigma_A = 0.29$ if
$A>1$ (Verde et al. 2003). The spectral index differs now from a 
scale invariant $n_s = 1$ primordial spectrum at the 2$\sigma$
level. The marginalized best fit value is $n_s = 0.955 \pm
0.023$.  The right panel shows result for the WMAP data combined 
with the DM power spectrum as inferred by VHS. For this data we use a
prior with $\sigma_A = 0.29$ throughout, which is twice the
uncertainty of the {\em rms} fluctuation amplitude as discussed in
Section \ref{data}.
As discussed in detail by VHS, the  inferred amplitude is about 20\% 
larger compared to that inferred by the C02 data with the lower 
effective optical depth. 
The spectral index is consistent with a scale invariant spectrum with 
$n_s = 0.992 \pm 0.032$.  There is {\em no} significant evidence for a
non-scale invariant spectrum in agreement with the simpler
analysis in VHS which obtained $\ns=1.01\pm 0.02 (\rm{stat.})\pm 0.06
(\rm{syst.})$ and $\sigma_8=0.93\pm 0.03 (\rm{stat.})\pm 0.09
(\rm{syst.})$.

\begin{figure*}
\center\resizebox{1.0\textwidth}{!}{\includegraphics{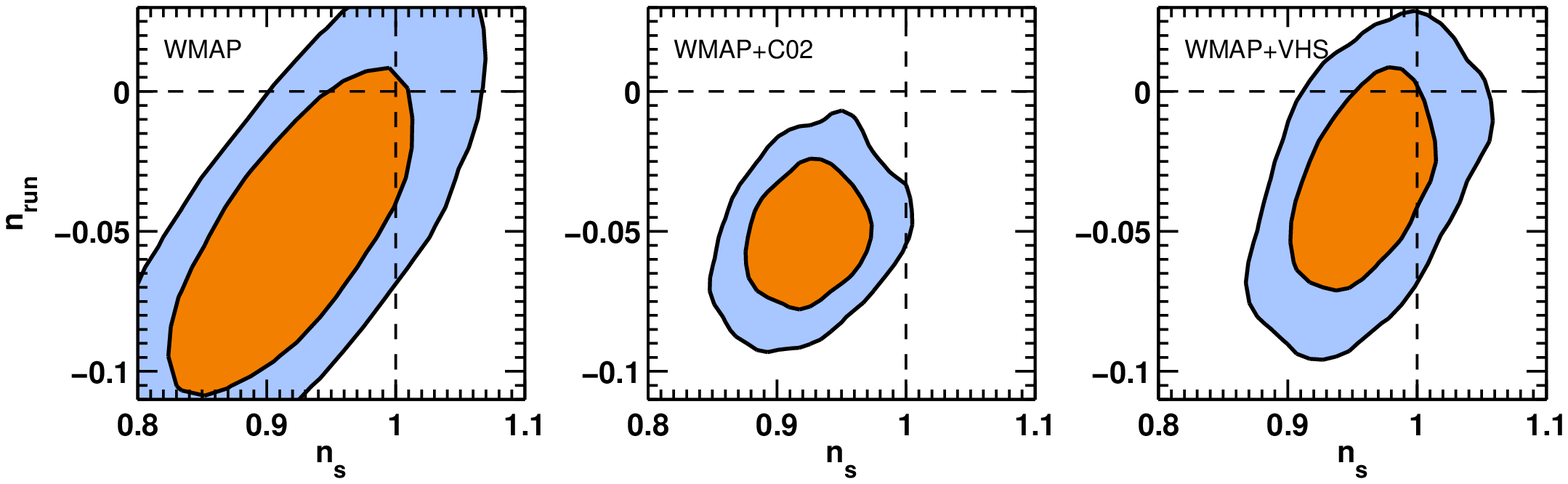}}
\caption{Joint $1$- and $2$-$\sigma$ likelihoods on $\nrun$ and
$\ns$. In the left panel the constraints are from the WMAP only, the middle
panel from WMAP combined with the \lya~power spectrum of C02, for a
$\teff (z=2.72)=0.349$. On the right the constraints from WMAP combined
with the results of VHS, for a $\teff (z=2.125) = 0.17$
and $\teff (z=2.72) = 0.305$.  The dashed lines indicate  a
scale-invariant $\ns=1$ and $\nrun=0$ primordial spectrum.}
\label{fig:run}
\end{figure*}

\begin{figure}
\center\resizebox{0.45\textwidth}{!}{\includegraphics{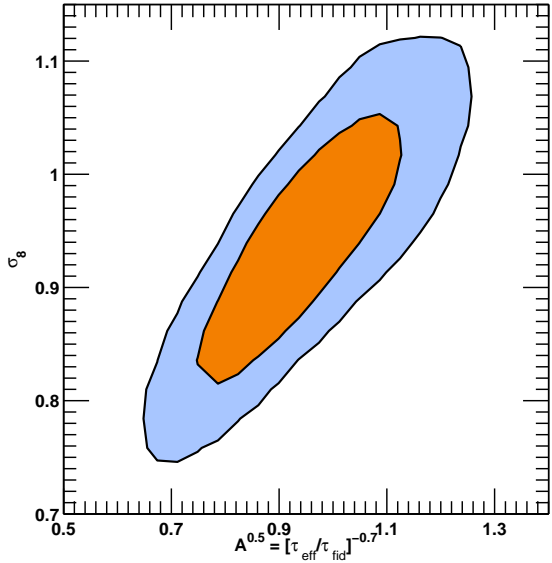}}
\caption{Joint $1$- and $2$-$\sigma$ likelihoods on $\sigma_8$ as a
function of the amplitude
of the power spectrum $A$ and of the effective optical depth.}
\label{fig:siga}
\end{figure}

In Figure~\ref{fig:siga}  we show the dependence of  the $\sigma_{8}$
inferred from the combined analysis on the amplitude of the power 
spectrum inferred from the \lya forest (parameterized 
as $A$, eq. 3). The dependence around the best fitting value is quite
close to linearity ($\sigma^2_8 \approx A)$. This plot 
was obtained with a prior on A given by eq. (3). The contours close at small and 
large values due to the constraints by the WMAP data alone. 
On the same axis we show  explicitly the dependence on the effective optical depth 
(relative to the assumed best fitting value).  The degeneracy 
of the inferred amplitude with the assumed effective optical depth 
already discussed by C02 and Seljak et al. (2003) is clearly seen.   
For the relation between amplitude A and effective optical depth 
we assumed $ A \propto \tau_{\rm eff}^{-1.4}$ as found by  VHS.

We will now discuss the constraints which we obtain if we include a running spectral
index.  In Figure \ref{fig:run} we show the joint $1\sigma$ and
$2\sigma$ likelihoods in the $n_{\rm run}$-$n_s$ plane marginalized
over all other parameters. In the left panel we show the constraints
from WMAP only. Again there is no significant evidence for a deviation
from a simple scale invariant power spectrum, although the marginalized
best fit values are $\ns = 0.922 \pm 0.062$ and $\nrun = -0.050 \pm
0.037$. In the middle we show the result from combining WMAP with the
C02 data.  The best fit values are $n_s = 0.924 \pm 0.030$ and $\nrun
= -0.052 \pm 0.016$.  This would rule
out the most simple scale invariant power spectrum on the $3$-$\sigma$
level. This is consistent with the findings of the WMAP team but
somewhat unusual for slow-roll inflationary models which require
$\nrun$ to be of the order of $(|\ns-1|)^2$ (Kosowsky \& Turner 1995;
but see Dodelson \& Stuart 2002).
If we instead  combine the VHS data with WMAP
(right panel) the contours shift again towards a scale invariant
spectrum with no running of the spectral index. The marginalized one
dimensional errors are $\ns = 0.959 \pm 0.035$ and $n_{\rm run} =
-0.033 \pm 0.025$. This makes the difference to a simple scale
invariant primordial power spectrum statistically insignificant.

\begin{table*}
\begin{center}
\caption{The marginalized results on cosmological parameters.
The quoted values correspond to the peaks
of the marginalized probability distributions. Errors are the 68\%
confidence limits. }
\label{tab:params}
\begin{tabular}{ccccc}
\hline
    &  WMAP+C02 (power law) &  WMAP+VHS (power law) &   WMAP+C02
    (running) &  WMAP+VHS (running) \\
\hline

$\Omega_{\rm c}h^2$ & $0.109 \pm 0.010$ & $0.124 \pm 0.013$ & $0.124 \pm 0.013$  & $0.132 \pm 0.015$ \\

$\Omega_{\rm b}h^2$ & $0.022\pm 0.009$ & $0.024\pm 0.001$ & $0.022 \pm 0.001 $ & $0.023 \pm 0.001$ \\

$h$ & $0.721\pm0.040$ & $0.704\pm0.047$ & $0.683 \pm 0.048$& $0.677 \pm 0.047$\\

$\tau$ & $0.115\pm 0.056$& $0.157 \pm 0.068$ & $0.192 \pm  0.065$& $0.182 \pm 0.069$ \\

$\sigma_8$ & $0.794 \pm 0.052$ & $0.936 \pm 0.079 $& $0.921 \pm 0.071$ &$0.978 \pm 0.090$ \\

$\ns$ & $0.955\pm 0.023$ & $0.992 \pm 0.032$ & $0.924 \pm 0.030$ &$0.959 \pm 0.036$\\

$n_{\rm{run}}$ & - & - & $-0.052\pm 0.016$& $-0.033 \pm 0.025$\\

\hline
\end{tabular}
\end{center}
\end{table*}
In Table \ref{tab:params} we show the constraints on all the
cosmological parameters for the combined WMAP and \lya analysis with
and without a running spectral index.

\subsection{Constraining Slow-Roll Inflation}\label{inflation}
In order to obtain model independent constraints on inflation we
concentrate on slow roll models. Since we are using a running of the
spectral index of the scalar perturbations over a wide range in
wavenumbers we use a second
order extension of the standard slow-roll approximation (Liddle \&
Lyth 1993). This is described by the parameters (Martin \& Schwarz
2000, Leach et al. 2002)
\beq
\epsilon_1 \simeq \frac{\Mp^2}{2}\left(\frac{V^\prime}{V}\right)^2\;,
\eeq
\beq
\epsilon_2 \simeq 2{\Mp^2}\left[\left(\frac{V^\prime}{V}\right)^2-\frac{V^{''}}{V}\right]\;,
\eeq
\beq
\epsilon_2\epsilon_3 \simeq
2{\Mp^4}\left[\frac{V^{\prime\prime\prime}V^{\prime}}{V^2}-3\frac{V^{\prime\prime}}{V}\left(\frac{V^\prime}{V}\right)^2+2\left(\frac{V^\prime}{V}\right)^4\right]\; ,
\eeq
where primes denote derivatives with respect to the field and
$\Mp=(8\pi G)^{-1/2}$ is the reduced Planck mass. Note that
$\epsilon_1, \,\epsilon_2,\,\epsilon_3$
correspond to the conventional slow-roll parameters by $\epsilon_1 =
\epsilon_V$, $\epsilon_2 = 2\epsilon_V-2\eta_V$ and $\epsilon_2
\epsilon_3 = 4 \epsilon_V^2 - 6\epsilon_V \eta_V +2\xi_V$.
If we calculate the spectrum of primordial perturbations in the
inflaton field we can identify the slow-roll parameters with
parameters of the power spectrum. In general inflation also predicts
tensor fluctuations or gravitational waves. We neglect here the
running of the tensor index which is given by $n_{\rm t}^{\rm run} =
\epsilon_1 \epsilon_2$. With the slow-roll conditions we obtain
\ba
\ns -1 & \approx& -2\epsilon_1 - \epsilon_2-2\epsilon_1^2-C\epsilon_2\epsilon_3\; ,\label{eqn:ns}\\
r &\approx& 16\epsilon_1 \, ,\label{eqn:r}\\
\nrun & \approx& -\epsilon_2\epsilon_3\, , \label{eqn:nr}
\ea
with $C \approx 0.73$. We then obtain as a ``consistency''
relation between the scalar to tensor ratio $r$ and the tensor
spectral index
\be
\nt = -\frac{r}{8}-\frac{r^2}{128}\; .
\label{eqn:nt}
\ee

Note that this is only true if we neglect the running of the tensor spectral index. We will
calculate the slow-roll quantities from the Markov chains generated in
the parameters $\ns$, $\nrun$ and $r$. In this
case we chose as the pivot scale $k_0 = 0.002\;{\rm Mpc}^{-1}$ (Peiris
et al. 2003).

We perform a parameter analysis as described in Section \ref{powerspectrum}
at this pivot scale with the addition of tensor modes with a free
amplitude ratio $r<1$ and the constraint in equation (\ref{eqn:nt}).
\begin{figure}
\center\resizebox{0.5\textwidth}{!}{\includegraphics{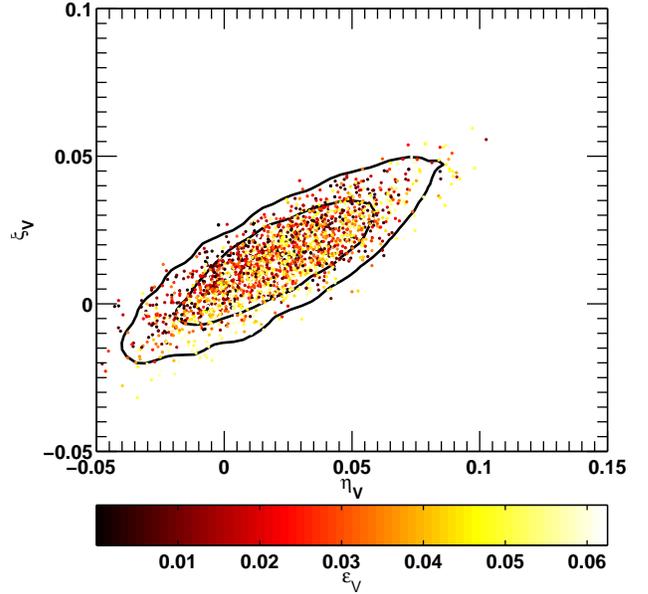}}
\caption{Joint likelihood in the $\eta_V$-$\xi_V$ plane, where the
  colour coding of the dots is scaled with $\epsilon_V$. The contours
 are the 68\% and 95\% confidence levels.}
\label{fig:slowroll}
\end{figure}
In Figure \ref{fig:slowroll} we show the results of this analysis. The
plot shows the results in the $\eta_V$-$\xi_V$ plane. The models are
colour coded
according to the third slow-roll parameter $\epsilon_V$. From
equation (\ref{eqn:r}) we see that $0<\epsilon_V<0.0625$, if we restrict
the analysis to $r<1$. For most models
of inflation there is a negligible contribution of gravitational waves
to the perturbations, i.e. $r \sim \epsilon_V \approx 0$ (Liddle \&
Lyth 2000). In this case
the likelihood contour in Figure \ref{fig:slowroll} is restricted to
the area depicted by the dark points, towards the lower left
of the figure. $\eta_V$ and $\xi_V$ are thus relatively small, which according to equations (\ref{eqn:ns}) and
(\ref{eqn:nr}) corresponds to a spectrum very close to scale invariant
scalar fluctuations. This is consistent with our findings
in Section \ref{powerspectrum}. However, with the current data, models
with a large tensor contribution are also valid, which correspond to
the light coloured points in Figure 4, in the top right corner. The
marginalized best fit values for the slow-roll parameters are
$\epsilon_V = 0.032 \pm 0.018$, $\eta_V = 0.020 \pm 0.025$ and $\xi_V =
0.015 \pm 0.014$. The constraint on
the tensor to scalar ratio is $r = 0.499 \pm 0.296 $.

\section{Conclusion}\label{conclu}
We have combined the linear dark matter power spectrum inferred from
the \lya forest with the  1st year WMAP  data to obtain constraints on 
the {\it rms} fluctuation amplitude of matter fluctuations $\sigma_{8}$
 and on the slope and shape parameter of the primordial density power
spectrum $\ns$ and $\nrun$.  We have restricted our analysis to
spatially flat models with a low gravitational wave contribution to the 
CMB fluctuations. The WMAP data alone provides only weak constraints on 
 $\sigma_{8}$, $\ns$ and $\nrun$. Adding the \lya forest significantly tightens the
constraints because of the resulting long  lever arm which spans 
more than 3 orders of magnitude in wavenumber. The WMAP data combined
with  the DM power spectrum inferred by Croft et al. (2002) from their 
\lya forest data requires a  tilt ($\ns <1$) and/or running of the
spectral index ($\nrun  <0$) to fit the
data which is significant at the 3 $\sigma$ level. 
If we combine instead the WMAP data with the DM power spectrum inferred by VHS
a smaller tilt and running of the spectral
index are required which are not statistically significant. 
The best fit value for the {\it rms} fluctuation amplitude
of matter fluctuations is $\sigma_{8} = 0.94 \pm 0.08 $  and $\ns =
0.99 \pm 0.03$, in good agreement 
with the simpler analysis presented by VHS. 

The difference of the analysis for the DM power spectrum of C02 and
VHS is mainly due to the more appropriate
smaller effective optical depth obtained from high resolution, high S/N
QSO absorption spectra which was assumed in VHS. We
further included tensor modes and put constrains on the slow roll
parameters. At present the limits on $\epsilon_V$, $\xi_V$ and
$\eta_V$ are not very tight, but future CMB missions such as Planck, together with
better understanding of the systematic uncertainties of the
Lyman-$\alpha$ forest data  will
improve the constraints.

\section*{Acknowledgements.} 
This work is supported by the European Community Research and Training
Network ``The Physics of the Intergalactic Medium''. The simulations
were done at the UK National Cosmology Supercomputer Center funded by 
PPARC, HEFCE and Silicon Graphics / Cray Research. We thank
Sarah Bridle, George Efstathiou, Sabino Matarrese, Hiranya Peiris for useful discussions, Anthony
Lewis for technical help and useful discussions and Dominik Schwarz 
for helpful comments on the manuscript.

\def\apj{Ap.\ J.}
\def\apjs{Ap.\ J. \ S.}

\def\mnras{MNRAS}
\def\mn{MNRAS}

\end{document}